\documentclass{article}
\usepackage[centertags]{amsmath}
\usepackage{amsfonts} \usepackage{amssymb} \usepackage{amsthm}
\allowdisplaybreaks[1]
\usepackage{graphicx}

\newcommand{\cM}{{\mathcal{M}}}
\newcommand{\cN}{{\mathcal{N}}}
\newcommand{\cO}{{\mathcal{O}}}

\newcommand{\one}{{\rm 1\kern -.9mm l}}

\newcommand{\ii}{\mathrm{i}}
\newcommand{\ee}{\mathrm{e}}

\newcommand{\vev}[1]{\langle #1 \rangle}

\newcommand{\tr}{\mathrm{tr}\,}

\newcommand{\href}[2]{#2}

\begin{document}
\def\thefootnote{\alph{footnote}}
\begin{center}
{\Large\bf 
Non-perturbative aspects of gauge/gravity duality%
\footnote{Proceedings of the Corfu Summer Institute 2012 \emph{``School and Workshops on Elementary 
Particle Physics and Gravity''}, September 8-27, 2012, Corfu, Greece. Luca Giacone was the speaker.}}
\end{center}
\vskip 0.5cm
\centerline{
Marco Bill\'o${}^1$, Marialuisa Frau${}^1$, Luca Giacone${}^{1}$ and Alberto Lerda${}^2$}
%
\vskip0.4cm
\centerline{\sl ${}^1$ Universit\`a di Torino, Dipartimento di Fisica and}
\centerline{\sl I.N.F.N. - Sezione di Torino}
\centerline{\sl via P.Giuria 1, I-10125 Torino, Italy}
\vskip 0.3cm
\centerline{\sl ${}^2$ Universit\`a del Piemonte Orientale,}
\centerline{\sl  Dipartimento di Scienze e Innovazione Tecnologica and}
\centerline{\sl I.N.F.N. - Gruppo Collegato di Alessandria - Sezione di Torino}
\centerline{\sl Viale T. Michel 11, I-15121 Alessandria, Italy.}
\vskip 0.3cm
\centerline{\sl e--mail: {\tt billo, frau, giacone, lerda@to.infn.it}}

\vskip 0.5cm
\begin{abstract}
Recently we provided a microscopic derivation of the exact supergravity
profile for the twisted scalar field emitted by systems of fractional D3-branes 
at a $\mathbb{Z}_2$ orbifold singularity. In this contribution we
focus on a set-up supporting
an $\mathcal{N} = 2$ SYM theory with SU$(2)$ gauge group and $N_f=4$.
We take into account the tower of D-instanton corrections to the source
terms for the twisted scalar
and find that its profile can be expressed in terms of the chiral ring elements of
the gauge theory.
We show how the twisted scalar, which at the perturbative level represents
the gravity counterpart of the gauge coupling, at the non-perturbative
level is related to the effective gauge coupling in an interestingly
modified way.
\end{abstract}
\setcounter{footnote}{0}
\def\thefootnote{\arabic{footnote}}


\section{Introduction}
\label{sec:intro}
The embedding of supersymmetric gauge theories in a string framework using systems of D-branes has 
been very fruitful and inspiring for many developments. For example,
the famous AdS/CFT correspondence \cite{Maldacena:1997re} is rooted
in the realization of the $\cN=4$ super Yang-Mills (SYM) theory by means of D3-branes in flat space 
and in the profile of the supergravity bulk fields they induce in space-time. 
In less supersymmetric and/or in non-conformal cases (like the $\cN=2$ gauge theories 
in four dimensions we will be interested in) the corresponding gravitational profile 
depends on some transverse directions representing the energy scale 
thus accounting for the running of the gauge theory. This fact was explicitly checked long ago \cite{Klebanov:1999rd}\nocite{Bertolini:2000dk,Polchinski:2000mx,Petrini:2001fk,Bertolini:2001qa}
-\cite{Billo:2001vg} at the perturbative level
in $\cN=2$ SYM theories realized by fractional D3 branes of type IIB at non-isolated singularities, like for instance the $\mathbb{C}^2/\mathbb{Z}_2\times\mathbb{C}$ orbifold.
By studying the emission of closed string fields from such branes, the corresponding ``perturbative'' supergravity solutions were constructed and it was found that
a scalar field from the twisted sector, which we will call $t$, varies logarithmically 
in the internal complex direction $z$ transverse to the orbifold, matching precisely the
perturbative logarithmic running of the gauge coupling with the energy scale. 
However, such perturbative solutions suffer from singularities at small values 
of $z$, {\it i.e.} in the IR region of the gauge theory, and have to be modified 
by non-perturbative corrections.

It is well-known that in $\cN=2$ gauge theories there is a whole series 
of non-perturbative contributions to the low-energy effective action that are due to instantons. 
In the last two decades tremendous advances 
have been made in the study of instanton effects within field theory (for reviews, 
see for instance \cite{Dorey:2002ik,Bianchi:2007ft}), and more recently
also within string theory by means of D-instantons, {\it i.e.} D-branes
with Dirichlet boundary conditions in all directions \cite{Witten:1995gx}\nocite{Douglas:1995bn,Green:2000ke}-\cite{Billo:2002hm}.
In the seminal papers \cite{Seiberg:1994rs,Seiberg:1994aj} the exact solutions 
for the low-energy effective $\cN=2$ theories in the Coulomb branch, including all instanton
corrections, were found using 
symmetry and duality arguments. In particular it was shown that
the effective SYM dynamics in the limit of low energy and momenta can be
exactly encoded in the so-called Seiberg-Witten (SW) curve which describes 
the geometry of the moduli space of the SYM vacua. Later these results were
rederived from a microscopic point of view with the help of localization techniques
\cite{Nekrasov:2002qd,Nekrasov:2003rj} that permit an explicit evaluation of the integrals 
over the multi-instanton moduli space. These techniques fit naturally in the string/D-brane context 
and indeed have been exploited for interesting generalizations of the SW
results in many different directions.

It is then natural to ask how the infinite tower of instanton effects is encoded in the dual holographic description of the gauge theory in terms of gravity. 
To answer this question one possibility is to exploit symmetry and duality arguments and 
determine the background geometry that incorporates the exact SW solution, like in the 
M-theory constructions based on configurations of D4 and NS5 branes \cite{Witten:1997sc}. 
Another possibility is to compute directly the multi-instanton corrections to the profiles 
of the gravitational bulk fields%
\footnote{Recently, a different, but related, interesting approach to the gravitational dual of a gauge theory through the study of the emergent geometry associated to the D-instanton effective action has been developed in \cite{Ferrari:2012nw}\nocite{Ferrari:2013pi}-\cite{Ferrari:2013wla}.}.
This is what we will discuss in this contribution, which 
heavily relies on the content of \cite{Billo:2011uc,Fucito:2011kb,Billo:2012xj} and especially of
\cite{Billo:2012st}.
In particular we will briefly review how to derive the exact supergravity profile of 
the twisted field $t$ emitted by a system of fractional 
D3-branes at a $\mathbb Z_2$-orbifold singularity supporting a $\cN=2$ quiver gauge theory 
with unitary groups and bi-fundamental matter, and show how to obtain from it the exact
running of the gauge coupling constant, including the non-perturbative contributions, in
perfect agreement with the SW solution. 

\section{The gauge coupling for the SU(2) $N_f=4$ SYM theory}
\label{secn:setup}

\begin{figure}
\begin{center}
\begin{picture}(0,0)%
\includegraphics{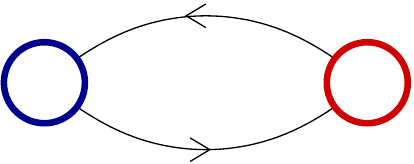}%
\end{picture}%
\setlength{\unitlength}{1657sp}%
\begingroup\makeatletter\ifx\SetFigFontNFSS\undefined%
\gdef\SetFigFontNFSS#1#2#3#4#5{%
  \reset@font\fontsize{#1}{#2pt}%
  \fontfamily{#3}\fontseries{#4}\fontshape{#5}%
  \selectfont}%
\fi\endgroup%
\begin{picture}(4706,1844)(1203,-3953)
\put(1441,-3121){\makebox(0,0)[lb]{\smash{{\SetFigFontNFSS{9}{10.8}{\rmdefault}{\mddefault}{\updefault}$N_0$}}}}
\put(5176,-3121){\makebox(0,0)[lb]{\smash{{\SetFigFontNFSS{9}{10.8}{\rmdefault}{\mddefault}{\updefault}$N_1$}}}}
\end{picture}%
\end{center}
\caption{The quiver diagram for the orbifold $\mathbb C^2/\mathbb Z_2$. The branes
of type 0 are represented by the blue circle while the branes of type 1 are represented
by the red circle. The oriented lines connecting the two types of branes represent 
the hypermultiplets in the bifundamental representations.}
\label{fig:quiver}
\end{figure}
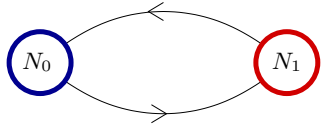

We study the prototypical case of $\mathcal N=2$ SYM theories that are realized with
fractional D3-branes at the non-isolated orbifold singularity 
$\mathbb C^2/\mathbb Z_2\times \mathbb C$. 
In this orbifold there are two types of fractional D3-branes, which we call types 0 and 1,
corresponding to the two different irreducible representations
of $\mathbb Z_2$.
The most general brane configuration therefore consists of $N_0$ branes of type 0 and $N_1$
branes of type 1, and corresponds to an ${\mathcal N}=2$ quiver theory in four dimensions
with gauge group U($N_0$)$\times$U($N_1$) and with a matter content given by one hypermultiplet
in the bi-fundamental representation $({\mathbf{\overline N_0}},{\mathbf{N_1}})$ 
and one hypermultiplet in the $({\mathbf{N_0}},{\mathbf{\overline N_1}})$ representation. 
The corresponding quiver diagram is
represented in Fig.~\ref{fig:quiver}.
Ignoring the gauge degrees of freedom on the $N_1$ branes, one obtains
an ${\mathcal N}=2$ U($N_0$) SYM theory with $2N_1$ fundamental flavors and U($N_1$) as
global symmetry group. Furthermore, we will decouple the U(1) factors and concentrate 
on the SU($N_0$)$\times$ SU($N_1$) part of the symmetry group. 

In this contribution we focus on the case $N_0=N_1=2$, 
representing an $\cN=2$ SU(2) SYM theory with $N_f=4$
flavors, but our results and methods apply to the general case as well \cite{Billo:2012st}. 
The SU(2) $N_f=4$ SYM theory has a vanishing $\beta$-function but, when the 
flavors are massive, the gauge coupling gets renormalized at 1-loop 
by terms proportional to the mass parameters of the hypermultiplets. 
This situation corresponds to placing the
fractional D3-branes at different positions in the transverse plane, {\it i.e.}
to giving non-vanishing vacuum expectation values to the adjoint scalars $\phi_0$ and $\phi_1$
of the vector multiplets on the two types of branes according to
\begin{equation}
\langle \phi_0 \rangle =
\mathrm{diag}(a,-a)~~~\mbox{and}~~~\langle \phi_1 \rangle  = \mathrm{diag}(m,-m)~.
 \label{am}
\end{equation}
Note that this brane configuration implies that the masses of the four flavors are given by
\begin{equation}
 \{m_1,m_2,m_3,m_4\}=\{m,-m,m,-m\}~.
\label{masses}
\end{equation}
In this case one finds that the perturbative part of gauge coupling constant 
\begin{equation}
 \tau \equiv \frac{\theta_{\mathrm{YM}}}{\pi}+\ii\frac{8\pi}{g_{\mathrm{YM}}^2}
\end{equation}
is given by
\begin{equation}
 \ii\pi  \tau_{\mathrm{pert}} =
\ii\pi  t_0 +\ii\pi-\log16+2\log\Big(1-\frac{m^2}{a^2}\Big)
\label{taupert}
\end{equation}
where $t_0$ is the bare coupling. Besides these perturbative terms, there are also non-perturbative corrections due to instantons which can be explicitly computed using localization techniques
(see for instance \cite{Billo:2012st,Billo:2013fi} for details). The first two instanton contributions
turn out to be given by
\begin{equation}
 \ii\pi  \tau_{\mathrm{inst}} =
q_0\Big(\frac{1}{2}+\frac{3m^4}{2a^4}\Big)+q_0^2\Big(\frac{13}{64}+\frac{33m^4}{32a^4}
-\frac{15m^6}{8a^6}+\frac{105m^8}{64a^8}\Big)+\cO(q_0^3)
\label{tauinst}
\end{equation}
where $q_0=\ee^{\pi\ii t_0}$ is the instanton counting parameter.
The complete effective coupling is therefore the sum of (\ref{taupert}) and (\ref{tauinst}).
For our future considerations it is convenient to rewrite it in terms of the gauge invariant quantity
\begin{equation}
 \mathbf{v} =  \frac{1}{2}\,\vev{\tr \phi_0^{2}}
\label{v}
\end{equation}
which parametrizes the moduli space of the effective theory at the quantum level. 
Using the multi-instanton calculus and localization techniques,
one can show that $\mathbf{v}$ is related to 
the classical vacuum expectation value $a$ in the following way \cite{Flume:2004rp,Billo:2012st}
\begin{equation}
 \mathbf{v} = a^2\Big(1+q_0\,\frac{(a^2-m^2)^2}{2a^4}
+q_0^2\,\frac{(a^2-m^2)(13a^6-15a^4m^2+7a^2m^4-5m^6)}{32a^8}+\cO(q_0^3)\Big)
\label{va}
\end{equation}
Inverting this relation and substituting it into (\ref{taupert}) and (\ref{tauinst}), after some
simple algebra we find
\begin{eqnarray}
\ii\pi\,\tau(\mathbf{v}) &=&
 \log q_0+\ii \pi  -\log 16+2\log\Big(1-\frac{m^2}{\mathbf{v}}\Big)
+\,q_0\Big(\frac{1}{2}- \frac{m^2}{\mathbf{v}} +  \frac{5m^4}{2\mathbf{v}^2}\Big)
\nonumber\\
&&\,+\,q_0^2\Big(\frac{13}{64}-\frac{13m^2}{16 \mathbf{v}}+\frac{135m^4}{32\mathbf{v}^2}
-\frac{109m^6}{16\mathbf{v}^3}+\frac{269m^8}{64\mathbf{v}^4}\Big)+\cO(q_0^3)~.
\label{taumv}
\end{eqnarray}
For simplicity, and also for later convenience, we have introduced a notation that 
explicitly exhibits only the dependence of $\tau$ on the gauge 
invariant parameter $\mathbf{v}$.

When the flavors are massless, the effective coupling, which we denote by $\tau_0$, 
is related to $q_0$ as follows 
\begin{equation}
\ii\pi \tau_0
=\log q_0 +\ii \pi -\log 16+\frac{1}{2}\,q_0+\frac{13}{64}\,q_0^2+\frac{23}{192}\,q_0^3
+\cdots~.
\label{tau0m}
\end{equation}
It is interesting to observe that the inverse relation can be expressed in terms 
of modular functions. Indeed, inverting (\ref{tau0m}) we obtain
\begin{equation}
q_0=-16 \,\big( \ee^{\ii\pi\tau_0} +8 \,\ee^{2\ii\pi \tau_0}+44\, \ee^{3\ii\pi \tau_0}
+ \cdots\big) = -\frac{\theta_2^4\big(\tau_0\big)}{\theta_4^4\big(\tau_0\big)}
\label{q0tau0}
\end{equation}
where the $\theta$'s are the Jacobi $\theta$-functions%
\footnote{Notice that this same relation can also be written as $q_0=-16\,\frac{\eta^8\big(4\tau_0\big)}{\eta^8\big(\tau_0\big)}$ in terms of the Dedekind $\eta$-functions. This expression is amenable of interesting generalizations for superconformal
field theories with gauge groups SU($N$) with $N>2$ \cite{Billo:2012st}. We also observe that
our coupling is related by a T-duality transformation, $\tau_0\to\tau_0+1$, to the one 
usually considered in the literature \cite{Grimm:2007tm,Alday:2009aq,Billo:2013fi} for which the relation (\ref{q0tau0}) takes the form $q_0=\frac{\theta_2^4\big(\tau_0\big)}{\theta_3^4\big(\tau_0\big)}$.}. Notice
that even in this simple case, $t_0$ and $\tau_0$ are different and represent two 
different choices of effective couplings for the massless theory (see also \cite{Billo:2010mg}).

Let us now consider the same modular function appearing in (\ref{q0tau0}), but instead of the massless coupling $\tau_0$ let us use as argument the massive coupling $\tau(\mathbf{v})$:
\begin{equation}
-\frac{\theta_2^4\big(\tau({\mathbf{v}})\big)}
{\theta_4^4\big(\tau({\mathbf{v}})\big)}\equiv q(\mathbf{v})
= -16 \,\Big( \ee^{\ii\pi\tau({\mathbf{v}})} 
+8 \,\ee^{2\ii\pi \tau({\mathbf{v}})}+44\, \ee^{3\ii\pi \tau({\mathbf{v}})}
+ \cdots\Big)~.
\label{qtheta}
\end{equation}
Taking the logarithm of $q({\mathbf{v}})$, we then get
\begin{eqnarray}
 \ii\pi\,t({\mathbf{v}}) &\equiv& \log q({\mathbf{v}})
\label{tmv}\\
&=&\log q_0+ 2\log\Big(1-\frac{m^2}{\mathbf{v}}\Big)
+\,q_0\,\frac{2m^4}{\mathbf{v}^2}+\,q_0^2
\Big(\frac{2m^4}{\mathbf{v}^2}
-\frac{4m^6}{\mathbf{v}^3}+\frac{3m^8}{\mathbf{v}^4}\Big)+\cO(q_0^3)~.
\nonumber
 \end{eqnarray}
This expression has a very nice interpretation. Indeed, let us consider the SW curve for the
SU(2) $N_f=4$ SYM which, when the flavor masses are as in (\ref{masses}), can be written as 
\cite{Argyres:1999ty,Billo:2012st}
\begin{equation}
 y^2=P^2(z)-g^2\,Q(z)
\label{curve}
\end{equation}
where
\begin{equation}
 P(z)=z^2-u~,\qquad Q(z)=(z^2-m^2)^2~,
\label{PQ}
\end{equation}
with
\begin{equation}
 {u}=\frac{1-q_0}{1+q_0}\,\mathbf{v}+\frac{2q_0}{1+q_0}m^2
~,\qquad g^2=\frac{4q_0}{(1+q_0)^2}~.
\label{ug2}
\end{equation}
The curve (\ref{curve}) describes a torus whose complex structure parameter
is
\begin{equation}
 \tau=\frac{\partial {a_D}}{\partial a}
\label{tauaad}
\end{equation}
where $a$ and $a_{D}$ are the periods of the SW 1-form differential $\lambda$
\begin{equation}
a=\frac{1}{2\ii\pi} \oint_{\gamma} \lambda~,~~~~
{a_D} = \frac{1}{2  \ii\pi}\, \oint_{\widetilde \gamma} \lambda
\label{aad}
\end{equation}
computed around a basis of dual cycles 
$(\gamma,\widetilde\gamma)$ normalized in such a way that $\gamma\circ \widetilde\gamma=1$.
The SW differential $\lambda$ can be written as
\begin{equation}
\lambda= z\, \Psi'(z)\, dz
\label{SWdifferentialnoi}
\end{equation}
with
\begin{equation}
\Psi(z) = \log \frac{P(z)+\sqrt{P^2(z) - g^2 Q(z)}}{\mu^2}~.
\label{psi}
\end{equation}
Using this information, one can compute $\tau$ and check that,
when it is expanded in powers of $q_0$, it coincides precisely with the effective gauge coupling (\ref{taumv}).
Other interesting quantities that characterize the curve (\ref{curve}) 
are the anharmonic ratios of the four roots of the
equation $y^2=0$. It is quite easy to see that these roots are
\begin{eqnarray}
 e_1&=&+\sqrt{\frac{u-g\,m^2}{1-g}}~,~~~e_2=-\sqrt{\frac{u+g\,m^2}{1+g}}~,\nonumber\\
e_3&=&-\sqrt{\frac{u-g\,m^2}{1-g}}~,~~~e_4=+\sqrt{\frac{u+g\,m^2}{1+g}}~,
\label{roots}
\end{eqnarray}
and that a corresponding anharmonic ratio is
\begin{equation}
  \zeta \equiv\frac{(e_3-e_2)(e_1-e_4)}{(e_1-e_2)(e_3-e_4)}
=\frac{\mathbf{v}-\frac{2q_0}{1+q_0}\,m^2-\sqrt{(\mathbf{v}-\frac{2q_0}{1+q_0}\,m^2)^2
-\frac{4q_0}{(1+q_0)^2}\,(\mathbf{v}-m^2)^2}}{\mathbf{v}-\frac{2q_0}{1+q_0}\,m^2
+\sqrt{(\mathbf{v}-\frac{2q_0}{1+q_0}\,m^2)^2-
\frac{4q_0}{(1+q_0)^2}\,(\mathbf{v}-m^2)^2}}~.
 \label{ratio1}
\end{equation}
It is easy to see that the expansion of $\log \zeta$ in powers of $q_0$ 
perfectly matches the expression found in
(\ref{tmv}) and thus we are led to the identification 
\begin{equation}
\ii\pi t({\mathbf{v}})=\log \zeta~.
\label{tlogz}
\end{equation}
This is not surprising since the relation between the anharmonic ratio $\zeta$ and
the complex structure parameter $\tau$ of a curve like (\ref{curve}) is precisely $\log\zeta=
-\frac{\theta_2^4(\tau)}{\theta_4^4(\tau)}$, namely the same relation between 
$t({\mathbf{v}})$ and $\tau({\mathbf{v}})$ implied by
(\ref{qtheta}) and (\ref{tmv}). 

We observe observe that the right hand side of (\ref{ratio1}) can be nicely written in terms 
of the polynomials $P$ and $Q$ of the SW curve describing the theory 
at the so-called ``enhan\c{c}on vacuum''
\cite{Billo:2012st}. This is the specific point of the quantum moduli space 
corresponding to $\mathbf{v}=0$ which describes a classical extended
brane configuration resembling that of the enhan\c{c}on ring \cite{Benini:2008ir}. 
In the enhan\c{c}on vacuum we therefore have 
$u=\frac{2q_0}{1+q_0}m^2$, and the polynomials $P$ and $Q$ become
\begin{equation}
 \widetilde P(z) = z^2-\frac{2q_0}{1+q_0}m^2~\qquad \widetilde Q(z)=(z^2-m^2)^2~.
\end{equation}
Then, from (\ref{ratio1}) and (\ref{tlogz}) it is easy to realize that
\begin{equation}
 \ii\pi\,t({\mathbf{v}})=\log\frac{\widetilde P(z)-\sqrt{\widetilde P^2(z)-g^2\,\widetilde Q(z)}}
{\widetilde P(z)+\sqrt{\widetilde P^2(z)-g^2\,\widetilde Q(z)}}\Bigg|_{z^2=\mathbf{v}}~.
\label{texact}
\end{equation}

Using the information encoded in the SW curve it is also possible to compute the exact
quantum correlators $\big\langle \tr\,\phi_0^\ell\big\rangle$
forming the chiral ring elements of the gauge theory. These correlators are in fact given by the 
integral
\begin{equation}
 \big\langle\tr\, \phi_0^\ell\big\rangle = 
\oint_{ \gamma} \frac{dw}{2  \ii\pi}~w^\ell\, \Psi'(w) ~.
\label{trj}
\end{equation}
or alternatively, they can be obtained by expanding the generating functional \cite{Billo:2012xj}
\begin{equation}
\Big\langle \tr \, \frac{1}{z-\phi_0 } \Big\rangle  =  
  \Psi'(z) ~.
  \label{genphij}
\end{equation}
Integrating (\ref{genphij}) with respect to $z$, it is easy to find 
\begin{equation}
\Big\langle \tr\,\log \frac{z- \phi_0}{\mu}\Big\rangle 
=\log \frac{P(z)+\sqrt{ P(z)^2 - g^2 Q(z)}}{\mu^2} -\log\big(1+\sqrt{1-g^2}\big)~,
\label{trphi}
\end{equation}
where the integration constant has been fixed in order to match the $\cO(z^0)$ terms in the
expansion for large $z$ in both sides. With some further straightforward algebra, we can 
rewrite the right hand side of (\ref{trphi}) in the following form
\begin{equation}
\Big\langle \tr\,\log \frac{z- \phi_0}{\mu}\Big\rangle 
=\frac{1}{2}\,\log \frac{P(z)+\sqrt{ P(z)^2 - g^2 Q(z)}}{P(z)-\sqrt{ P(z)^2 - g^2 Q(z)}}
+\frac{1}{2}\,\log \frac{Q(z)}{\mu^{4}} + \frac{1}{2}\,\log q_0~.
\label{trphi1}
\end{equation}
This expression will be essential in the next section to write the exact 
({\it i.e.} all order in the instanton expansion) gravitational profile of the 
twisted scalar field $t$ emitted by the system
of fractional D3 branes, and to relate it with the dual gauge theory coupling.

\section{The exact $t$ profile emitted by fractional D3-branes and D-instantons}
The fractional D3-branes in the $\mathbb Z_2$ orbifold are gravitational sources
for a non-trivial metric and a 4-form R-R potential from the untwisted sectors,
and for two scalars, $b$ and $c$, from the twisted NS-NS and R-R sectors respectively
(see for instance \cite{Bertolini:2000dk}). While the
emitted untwisted fields can propagate in all six directions transverse to the D3-branes, 
the twisted scalars only propagate in the complex plane transverse to the D3-brane world-volume
which is not affected by the orbifold projection and which we parametrize with
a complex coordinate ${\mathbf x}$. A system of fractional D3-branes distributed 
on this plane therefore generate a non-trivial dependence of the fields $b$ and $c$ on ${\mathbf x}$.
The twisted scalars are conveniently combined in a complex field
\begin{equation}
 t= c +\tau\,b
\label{t}
\end{equation}
where here $\tau$ stands for 
the axio-dilaton of the type IIB string theory. For simplicity we assume that
the axion is trivial and that there are no branes other than the fractional D3 branes
so that the dilaton does not run. Thus, in this case we simply have $\tau=\ii/g_s$ where
$g_s$ is the string coupling constant. 
The field $t$ is actually part of a chiral bulk superfield $T$
whose structure is schematically given by
\begin{equation}
 T=t+\cdots+\theta^4\, \frac{\partial^2 }{\partial {\bf x}^2} \, \bar t+\cdots
\label{T}
\end{equation}
with dots denoting the supersymmetric descendants of $t$ and $\bar t$ being 
the complex conjugate of $t$. 

The profile of the twisted scalar $t$ emitted by a system of fractional D3-branes
can be derived by solving the classical field equations that
follow from the bulk action containing the kinetic terms and the source action
describing the emission from the fractional D3-branes. At the perturbative level 
this profile was obtained long ago in Ref.s~\cite{Bertolini:2000dk}\nocite{Polchinski:2000mx,Petrini:2001fk,Bertolini:2001qa}
-\cite{Billo:2001vg} and for a system of $N_0$ branes of type 0 and $N_1$ branes of type 1
located at the origin is
\begin{equation}
 \ii\pi t = \ii\pi t_0-2(N_0-N_1) \log\frac{ {\mathbf x}   }{ {\mathbf x} _0}
\label{proft}
\end{equation}
where $t_0= \ii/(2g_s)$ and ${\mathbf x}_0$ is an arbitrary length scale. 
It is convenient to introduce the quantities
\begin{equation}
 z =\frac{  {\bf x}   }{2\pi\alpha'}~~~~\mbox{and}~~~~\mu=\frac{  {\bf x} _0}{2\pi\alpha'}
\label{ymu}
\end{equation}
with mass dimension 1, and rewrite the solution (\ref{proft}) as follows
\begin{equation}
 \ii\pi t = \ii\pi t_0-2(N_0-N_1) \log\frac{ z }{ \mu }~.
\label{proft1}
\end{equation}
Note that in the conformal cases ($N_0=N_1$), we simply have
\begin{equation}
 t = t_0 ~.
\label{proftconf}
\end{equation}

Let us now consider a more general configuration in which the D3-branes are not all at the
origin. This amounts to giving the adjoint scalars non-vanishing vacuum expectation values
as in (\ref{am}) (from now on we focus again only on the case $N_0=N_1=2$). Then, 
one can show that the $t$ profile corresponding to such a configuration is
 \begin{equation}
  \ii\pi t
           =\ii\pi t_0-2\,\tr\log \frac{z-\langle \phi_0\rangle}{\mu}
+2\,\tr\log \frac{z-\langle \phi_1\rangle}{\mu}
=\ii\pi t_0+2\log \frac{z^2-m^2}{z^2-a^2}~.
\label{proftam}
\end{equation}
It is not difficult to realize that this $t$ field satisfies
the following differential equation
\begin{equation}
\square\,t= 8 J_{\mathrm{cl}}\,\delta^2(z)
\label{tcl00}
\end{equation}
with
\begin{equation}
 J_{\mathrm{cl}}= \sum_{\ell=1}^\infty
\frac{\ii}{\ell\,!}\,\Big(\tr\langle \phi_0\rangle^{\ell} - \tr\langle \phi_1 \rangle^{\ell}\Big)
 \,\frac{\partial^{\ell}}{\partial z^{\ell}}
~=~ \ii\,\tr\ee^{\ii\,\bar p\,\langle\phi_0\rangle}
- \ii\,\tr\ee^{\ii\,\bar p\,\langle \phi_1\rangle}
 \label{tcl0}
 \end{equation}
where in the second step we introduced the momentum operator conjugate to $z$, that is
$\bar p = -\ii\partial/\partial z$.

The current $J_{\mathrm{cl}}$ has a nice interpretation in terms of disk diagrams describing
the couplings among the closed string twisted fields and the massless open string
excitations of the fractional D3-branes. Indeed, by considering the interactions of 
the NS-NS scalar $b$ (whose vertex operator we denote by $V_b$) with the scalar
$\phi_0$ (whose vertex we denote by $V_{\phi_0}$), we find
\begin{equation}
\sum_{\ell=0}^\infty \frac{1}{\ell\,!}\,\big\langle
\underbrace{V_{\phi_0}\cdots V_{\phi_0}}_{\ell}\,V_b\,\big\rangle_{\mathrm{D3}_0}
~=~ \frac{\pi}{g_s}\sum_{\ell=0}^\infty \frac{1}{\ell\,!}\,\tr\langle{\phi_0}\rangle^\ell\,(\ii\bar p)^\ell\,b
~=~\frac{\pi}{g_s}\,\tr\ee^{\ii\,\bar p\,\langle\phi_0\rangle}\,b~.
\label{Aell}
\end{equation}
This result follows by computing the correlation functions of the vertex operators using
standard CFT techniques as discussed for example in~\cite{Billo:2011uc}
and by frozing the scalars to their vacuum expectation values.
A completely similar calculation can be performed with the scalar $\phi_1$ of the type 1 branes
leading to
\begin{equation}
 -\frac{\pi}{g_s}\,\tr\ee^{\ii\,\bar p\,\langle \phi_1\rangle}b\,~.
\end{equation}
where the extra sign comes from the fact that branes of type 1 have opposite $b$-charge with
respect of those of type 0.
Adding an analogous term describing the interactions of the R-R twisted scalar $c$, we can write the total contribution to the effective action as
\begin{equation}
 -\ii\pi\Big(
 \tr\ee^{\ii\,\bar p\,\langle\phi_0\rangle}-
\tr\ee^{\ii\,\bar p\,\langle \phi_1\rangle}
\Big)\bar t~.
\label{intert}
\end{equation}
Supersymmetry requires that this interaction must be accompanied by other
structures (that could also be computed from string diagrams with extra fermionic insertions)
in such a way that the effective action follows from a holomorphic
prepotential. As discussed in~\cite{Billo:2011uc}\nocite{Fucito:2011kb,Billo:2012xj}-\cite{Billo:2012st} such a prepotential
is obtained simply by promoting the bulk and boundary scalars
to the corresponding chiral superfields. Denoting by $\delta T$ the fluctuation part of
$T$, one finds in particular the following term
\begin{equation}
\delta F_{\mathrm{cl}} = \ii\pi\Big(
\tr\ee^{\ii\,\bar p\,\Phi_0}-\tr\ee^{\ii\,\bar p\, \Phi_1}
\Big)\frac{\delta T}{\bar p^2}+\cdots
\label{F}
\end{equation}
where the dots represent interactions of higher orders in $\delta T$.
The effective action follows upon integrating the prepotential over $d^4\theta$;
when all four $\theta$'s are taken from $\delta T$ and the superfields $\Phi_0$ and
$\Phi_1$ are frozen to their vacuum expectation values, we recover precisely the interaction
(\ref{intert}). The classical current (\ref{tcl0}) therefore is associated to a source 
term for $t$ and is related to the prepotential (\ref{F}) in the following way
\begin{equation}
 J_{\mathrm{cl}}= \frac{\bar p^2}{\pi}\,
\frac{\delta F_{\mathrm{cl}}}{ \delta T}\Bigg|_{\Phi\to \langle \Phi \rangle}~.
\label{jclprep}
\end{equation}

Let us now investigate how the classical profile (\ref{proftam})
changes when non-perturbative effects due to gauge instantons are taken into account.
In our brane set-up, instantons are introduced by adding fractional D(--1)-branes. Since
we neglect the dynamics on the branes of type 1, we only consider the effects produced by
adding $k$ D-instantons of type 0.
The physical excitations of the open strings with at least one end-point on the
D(--1)-branes account for the instanton moduli which we
collectively denote as $\cM_{k}$.
They consist of the neutral sector, corresponding to D(--1)/D(--1) open strings
that do not transform under the gauge group,
and of the charged and flavored sectors arising respectively from the D(--1)/D3$_0$ and
D(--1)/D3$_1$ open strings. The complete list of instanton moduli and 
their transformation properties can be found in
\cite{Billo:2012xj,Billo:2012st}. 
Here we just recall that among the neutral moduli we have the
bosonic and fermionic Goldstone modes of the supertranslations of the
D3-brane world-volume which are broken by the D-instantons and which are
identified with the superspace coordinates $x$ and
$\theta$, and a complex scalar $\chi$ 
transforming in the adjoint representation of the instanton symmetry group 
U($k$), whose eigenvalues describe the position
of the D-instantons in the un-orbifolded directions transverse to the fractional D3-branes.

In order to find the non-perturbative $t$ profile we first compute the instanton induced prepotential
$F_{\mathrm{n.p.}}$ from which the non-perturbative source current $J_{\mathrm{n.p.}}$ can be
derived following a procedure similar to the one outlined for the classical current $J_{\mathrm{cl}}$.
The non-perturbative prepotential is defined as
\begin{equation}
 F_{\mathrm{n.p.}}= \sum_k \int \!\!d\widehat{\mathcal M}_{k}
~\ee^{-S_{\mathrm{inst}}({\mathcal M}_{k}, \Phi, T)}
\label{Fnp}
\end{equation}
where the integral is performed over the centered moduli $\widehat{\mathcal M}_{k}$, which include
all moduli except the superspace coordinates $x$ and $\theta$. Here
$S_{\mathrm{inst}}({\mathcal M}_{k}, \Phi, T)$ is the instanton action, describing
the interactions of the instanton moduli with the boundary and bulk superfields.
As explained in \cite{Billo:2012xj}, such an action is
\begin{equation}
S_{\mathrm{inst}}(\cM_{k}, \Phi, T) = -k\,\ii\pi\,t_0+ S'_{\mathrm{inst}}(\cM_{k}, \Phi)
-\ii\pi~\tr_{k}\ee^{\ii\bar p \,\chi}\,\delta T+\ldots
\label{sinst1}
\end{equation}
where $S'_{\mathrm{inst}}$ is the part accounting for the interactions
of the moduli among themselves and with the fields in the vector multiplet, and $\tr_k$ means
trace over the U($k$) indices. Inserting (\ref{sinst1}) in (\ref{Fnp}), to linear order 
in $\delta T$ we find
\begin{equation}
 \delta F_{\mathrm{n.p.}} = \ii\pi\,\delta T \, \sum_k q_0^{k}
\int \!\!d\widehat{\mathcal M}_{k}
~\ee^{-S'_{\mathrm{inst}}({\mathcal M}_{k}, \Phi)}~\tr_{k}\ee^{\ii\bar p \,\chi}~.
\label{Fnp2}
\end{equation}
The integration over the moduli space can be explicitly performed
using localization techniques and Nekrasov's approach to the multi-instanton calculus~\cite{Nekrasov:2002qd,Nekrasov:2003rj}. This amounts to
first define the deformed instanton partition function 
\begin{equation}
 Z_{\mathrm{inst}}= \sum_k q_0^{k}
  \int \!\!d {\mathcal M}_{k}
~\ee^{-S'_{\mathrm{inst}}({\mathcal M}_{k}, \Phi; \epsilon_1,\epsilon_2)}  
\label{Zinst}
\end{equation}
where $\epsilon_1$ and $\epsilon_2$ are deformation parameters which in our string set-up
can be introduced by putting the brane system in a graviphoton 
background~\cite{Billo:2006jm,Ito:2010vx}, and then to 
compute the prepotential according to
\begin{equation}
 F_{\mathrm{n.p.}}= -\lim_{\epsilon_1,\epsilon_2 \to 0}
\epsilon_1 \epsilon_2  \log  Z_{\mathrm{inst}}~.
\label{prepotinst}
\end{equation}
The integral appearing in (\ref{Fnp2}) is related to the instanton part of the chiral ring
elements $\big\langle \tr\,\phi_0^\ell\big\rangle$ of the gauge theory on the D3-branes, which
can be computed as%
\footnote{For details on the derivation of this formula we refer to \cite{Billo:2011uc,Fucito:2011kb}.}
\begin{equation}
\frac{1}{\ell!}\,\Big\langle
\tr\phi_0^{\ell} \Big\rangle_{\!\mathrm{inst}} =-\frac{1}{(\ell-2)\,!}\,
\lim_{\epsilon_1,\epsilon_2\to 0}\,
\frac{\epsilon_1 \epsilon_2}{Z_{\mathrm{inst}}} 
\sum_k \,q_0^{k}\int \!\!d {\mathcal M}_{k}
~\ee^{-S'_{\mathrm{inst}}({\mathcal M}_{k}, \Phi;\epsilon_1,\epsilon_2)}\,
\tr_{k}\chi^{\ell-2}~.
\label{trphi0}
\end{equation}
Notice that the integrals in (\ref{Zinst}) and (\ref{trphi0}) are over all
moduli including $x$ and $\theta$, and that in the limit $\epsilon_i \to 0$ the factor $\epsilon_1 \epsilon_2$ in (\ref{trphi0}) compensates for the volume $V\sim \frac{1}{\epsilon_1\epsilon_2}$ of the regularized four dimensional superspace.
Plugging (\ref{trphi0}) into (\ref{Fnp2}) one gets
\begin{equation}
 \delta F_{\mathrm{n.p.}}  = \ii\pi\,\Big\langle
\tr\ee^{\ii\,\bar p\,\Phi_0}
\Big\rangle_{\!\mathrm{inst}} ~ \frac{\delta T}{\bar p^2}
\label{Fnp3}
\end{equation}
which is nothing but the instanton completion of (\ref{F}). 

Adding the classical and the instanton contributions we obtain the full source current 
for $t$:
\begin{equation}
 J= \frac{\bar p^2}{\pi}\,\frac{\delta F}{ \delta T}\Bigg|_{\Phi\to \langle \Phi \rangle}
\!\!\!=\ii\,\Big\langle \tr\ee^{\ii\,\bar p\,\phi_0}\Big\rangle
 -\tr\ee^{\ii\,\bar p\,\langle \phi_1\rangle}
\label{Jex}
\end{equation}
where $\delta F=\delta F_{\mathrm{cl}}+\delta F_{\mathrm{n.p.}}$.
The field equation satisfied by $t$ is therefore
\begin{equation}
 \square\, t= 8 J\,\delta^2(z)~=~8
\sum_{\ell=0}^\infty
\frac{\ii}{\ell\,!}\,\Big[\big\langle \tr\phi_0^{\ell} \big\rangle- \tr\langle \phi_1\rangle^{\ell}    \Big]\,\frac{\partial^{\ell}}{\partial z^{\ell}}\,\delta^2(z)
\label{fecom}
\end{equation}
which is solved by
\begin{eqnarray}
 \ii\pi  t &=& \ii\pi  t_0 -2\,
\Big\langle \tr\log \frac{z- \phi_0}{\mu} \Big\rangle
+2\, \tr\log \frac{z- \langle\phi_1\rangle}{\mu} \nonumber\\
&=&
\ii\pi  t_0 -2\,
\Big\langle \tr\log \frac{z- \phi_0}{\mu} \Big\rangle
+2\log \frac{(z^2-m^2)}{\mu^2} ~.
\label{tauz}
\end{eqnarray}
This explicit solution shows that all non-trivial information about the $t$ profile
is contained in the ring of chiral correlators of the gauge theory defined on the D3-branes. 
This chiral ring accounts therefore for the full tower of D-instanton corrections 
to the gravity solution.

The chiral correlators $\big\langle \tr\,\phi_0^\ell\big\rangle$ can be computed from
(\ref{trphi0}) using Nekrasov's approach to the multi-instanton calculus. 
Equivalently (and more efficiently), as we have explained in the previous section,
they can be obtained from the SW curve describing the SYM theory.
In fact, inserting (\ref{trphi1}) in (\ref{tauz}) and taking into account the explicit definition
of $Q$ given in (\ref{PQ}), we can obtain the exact expression for the twisted scalar 
field emitted by the brane system, namely
\begin{equation}
 \ii\pi  t= \log
 \frac{P(z)-\sqrt{ P^2(z) - g^2 Q(z)}}{P(z)+\sqrt{P^2(z) - g^2 Q(z)}}~.
\label{tsugra}
\end{equation}
Our result generalizes the one derived in~\cite{Cremonesi:2009hq} for the pure SU($N$) SYM theories using supergravity and M-theory considerations, and is also perfectly consistent
with the findings of~\cite{Witten:1997sc} 
where the SU($N)$ SYM theory is realized in type IIA using D4 branes stretched between two
NS branes.

We can therefore say that the methods we have developed provide a microscopic derivation of the 
supergravity profile for $t$ in which a direct relation with the chiral ring elements of the
gauge theory on the source branes is clearly established and the non-perturbative effects
are explicitly explained in terms of fractional D-instantons.

\section{Conclusions}
We have considered a fractional D3-brane system in a $\mathbb{Z}_2$ orbifold supporting an $\mathcal{N}=2$ SYM theory with SU$(2)$ gauge group and $N_f=4$ flavors. 
We have considered the scalar field $t$ from the twisted closed string sector emitted by such a configuration, which, at the tree level, plays the r\^ole of the gauge coupling on the D3-branes.
As it is well known, the fractional D3-branes act as sources for $t$, so that $t$ has a logarithmic profile in the complex direction $z$ transverse to the orbifold; this profile matches the perturbative running of the gauge coupling if the transverse space is identified with the
Coulomb branch of the gauge theory.
We have taken into account the non-perturbative effects corresponding to the inclusion of (fractional) D-instantons and explicitly shown how they modify the source for $t$ and hence its profile. 
The moduli space integrals that determine the non-perturbative source terms are related to the ones appearing in the computation of the chiral ring operators of the gauge theory.
Through this relation, we can then express the profile of the twisted scalar as the quantum expectation value of its perturbative expression (see (\ref{tauz})). This, in turn, can be written 
in terms of the SW curve that describes the effective dynamics of the gauge theory on the Coulomb moduli space (see (\ref{tsugra})).

At the non-perturbative level, the gauge/gravity relation is deeply modified with respect to its perturbative standing. The twisted scalar $t$ can no longer be simply identified with the effective gauge coupling. However, if we consider the situation in which the source D3-branes sit at the ``enhan\c{c}on'' vacuum, $\mathbf{v}=0$, the scalar $t(z)$ is still directly, albeit non-trivially, related to the effective coupling $\tau(\mathbf{v})$ when $z^2$ is identified with the 
quantum Coulomb space variable $\mathbf{v}$. Indeed, in this case $t(z)$ is given by 
(\ref{texact}); this expression corresponds, according to (\ref{tlogz}), to the 
logarithm of the anharmonic ratio $\zeta$ which parametrizes the SW torus. The anharmonic ratio
is related to the complex structure $\tau$, namely to the effective gauge coupling, through the modular function appearing in (\ref{qtheta}).

In these proceedings we focused on the conformal SU(2) case, but in \cite{Billo:2012st} we showed
that a similar pattern occurs for higher rank conformal gauge theories, and also, after decoupling some flavors, for asymptotically free cases: the twisted scalar emitted by the branes at the enhan\c{c}on vacuum is related in the gauge/gravity correspondence to the low energy effective couplings via non-trivial modular functions which are generalizations of that appearing in (\ref{qtheta}).

\vskip 0.8cm
\noindent {\large {\bf Acknowledgments}}
\vskip 0.2cm
We warmly thank our coauthors of \cite{Billo:2012st}, Francesco Fucito, Francisco Morales and Daniel Ricci-Pacifici, for the fruitful and pleasant collaboration.
This work is partially supported by the MIUR-PRIN contract 2009-KHZKRX and by INFN through the projects MI12 and TV12.

\end{document}